\documentclass{webofc}
\usepackage[varg]{txfonts}   
\usepackage{amsmath}
\usepackage{amsfonts}
\usepackage{amssymb}
\usepackage{hyperref}
\usepackage{color}
\usepackage{graphicx}
\usepackage{epstopdf}
\usepackage{physics}
\def\@preprint{\@empty}
\newcommand\preprint[1]{\gdef\@preprint{\hfill #1}}
\usepackage{siunitx}
\usepackage{verbatim}
\usepackage{amsmath}
\usepackage{amsfonts}
\usepackage{multirow}
\usepackage{array,mathtools,booktabs}

\newcolumntype{C}{>{$}c<{$}}
\usepackage{cancel}
\usepackage{braket}
\usepackage{hyperref}

\usepackage{fancyvrb}
\usepackage{subfig}

\begin{document}
\title{Far from equilibrium Chiral Magnetic Effect in Strong Magnetic Fields from Holography}
%
%

\author{\firstname{Sebastian} \lastname{Grieninger}\inst{1,2}\fnsep\thanks{\email{sebastian.grieninger@uam.es}} \and
        \firstname{Sergio} \lastname{Morales-Tejera}\inst{1,2}\fnsep\thanks{\email{sergio.moralest@uam.es}} 
}

\institute{Instituto de F\'isica Te\'orica UAM/CSIC, c/Nicol\'as Cabrera 13-15, Universidad Aut\'onoma de Madrid, Cantoblanco, 28049 Madrid, Spain
\and
 Departamento de F\'isica Te\'orica, Universidad Aut{\'o}noma de Madrid, Campus de Cantoblanco, 28049 Madrid, Spain
          }

\abstract{%
We study the real time evolution of the chiral magnetic effect out-of-equilibrium in
strongly coupled anomalous field theories. 
We match the parameters of our model to QCD parameters and draw lessons of possible relevance for the realization of the chiral magnetic effect in heavy ion collisions. In particular, we find an equilibration time of about $\sim0.35$ fm/c in presence of the chiral anomaly for plasma temperatures of order $T\sim300-400$ M\text{eV}. 
\newline\vspace{0.15cm}
\small \textit{Preprint number:} IFT-UAM/CSIC-21-133}
\maketitle
\vspace{-0.75cm}
\section{Introduction}\vspace{-0.1cm}
\label{intro}
The chiral magnetic effect (CME) is the generation of an electric current in a chirally imbalanced medium by an applied magnetic field \cite{Vilenkin:1980fu,Fukushima:2008xe}.\footnote{For reviews see \cite{Kharzeev:2013ffa,Landsteiner:2016led,Kharzeev:2020jxw}.} Formally it is described by $\vec J = 8 c \mu_5 \vec B\,,$ where $c$ is the coefficient of the axial anomaly and $\mu_5$ the axial chemical potential. An experimental observation of the CME in the context of QCD may be possible in heavy-ion collisions. Recent experimental results of the collision of the isotopes ZrZr and RuRu~\cite{STAR:2021mii} highlight that there are many open questions in our understanding of the underlying physics.

The CME current is formally expressed in terms of a chemical potential (an equilibrium quantity) thus it is specially interesting to understand how the CME is realized far from equilibrium.  
Answering this question is particularly important in view of a possible realization of the CME in heavy ion collisions. In heavy ion collisions, the magnetic field is only present in the initial stages after the collision and decays quickly whereas the hot QCD matter has a short but non-vanishing equilibration time. The question naturally arising is if the CME builds up fast enough to be measurable before the magnetic field decays. We want to address this question in this work by modeling the strongly coupled quark gluon plasma by means of a holographic model. Previous related studies of out-of-equilibrium chiral transport in holography are \cite{Lin:2013sga,Ammon:2016fru,Grieninger:2017jxz,Haack:2018ztx,Cartwright:2020qov,Kharzeev:2020kgc,Landsteiner:2017lwm,Fernandez-Pendas:2019rkh,Morales-Tejera:2020xuv}. In this study~\cite{Ghosh:2021naw}, we investigate the realtime dynamics of the chiral magnetic effect in strong magnetic fields within holography for the first time in the full backreacted setup.

\section{Holographic model}
\label{sec-1}
We  study a holographic quantum field theory with a $U(1)_A\times U(1)_V$ symmetry which is manifest in the gravity theory in terms of an axial gauge field $A^{\mu}$ and a vector $V^{\mu}$ gauge field, respectively, with the associated field strengths denoted as $F_5=d A$ and $F=dV$. The anomaly is implemented through a Chern--Simons term which is gauge invariant up to a total derivative. We work with the \textit{consistent} form of the anomaly. Combining these ingredients, the holographic model we consider is the following\footnote{In our notation, Greek letters denote the bulk coordinates and small Latin letters the boundary coordinates.}
\begin{align}
\!\!S\!=\!\frac{1}{2\kappa^2}\!\!\int_{\mathcal{M}}\!\! \!d^5x\sqrt{-g}\left[R\!+\!\frac{12}{L^2}\!-\!\frac{1}{4}F^2\!-\frac{1}{4}F_{(5)}^2 \right. \left.\!\!\! +\frac{\alpha}{3} \epsilon^{\mu\nu\rho\sigma\tau} A_\mu\left( 3 F_{\nu\rho}F_{\sigma\tau}+F^{(5)}_{\nu\rho}F^{(5)}_{\sigma\tau}\right) \right]\! 
+\!S_{\!\text{GHY}}+\!S_{\!\text{ct}} \label{action} 
\end{align}
where $S_\text{GHY}$ is the Gibbons-Hawking-York boundary term to make the variational problem well defined, $L$ is the AdS radius, $\kappa^2$ is the Newton constant and $\alpha$ the Chern-Simons coupling. We supplement the action by the appropriate counter-terms $S_\text{ct}$ to cancel the divergences \cite{deHaro:2000vlm,Ammon:2020rvg}. The Levi-Civita tensor is defined as $\epsilon^{\mu \nu \rho \sigma \tau}=\epsilon(\mu \nu \rho \sigma \tau)/\sqrt{-g}\,$.  
From the action \eqref{action}, we may derive the equations of motion yielding 
\begin{align}
& \nabla_\nu F^{\nu\mu}+2\alpha \epsilon^{\mu\nu\rho\sigma\tau} F_{\nu\rho}F^{(5)}_{\sigma\tau}=0,\quad\quad \nabla_\nu F_{(5)}^{\nu\mu}+\alpha  \epsilon^{\mu\nu\rho\sigma\tau} \left( F_{\nu\rho}F_{\sigma\tau}+F_{\nu\rho}^{(5)}F_{\sigma\tau}^{(5)} \right)=0, \label{eom:axial}\\ 
& G_{\mu\nu}-\frac{6}{L^2} g_{\mu\nu}-\frac{1}{2} F_{\mu\rho}F_{\nu}^{\ \rho }
 +\frac{1}{8} F^2 g_{\mu\nu} -\frac{1}{2} F^{(5)}_{\mu\rho}F_{\nu}^{(5) \rho }
 +\frac{1}{8} F_{(5)}^{2}g_{\mu\nu}=0. \label{eom:Einstein}
\end{align}
We are interested in monitoring the vector current and pressure in the dual field theory. Both one-point functions can be extracted from the full bulk solution through the standard holographic prescription, i.e. varying the renormalized on-shell action with respect to the associated source term.
In infalling Eddington-Finkelstein coordinates\footnote{Note that $\xi$ parametrizes the anisotropy of the system, which is induced by the presence of a magnetic field $B$ which breaks the rotational invariance from SO(3) to SO(2) and points in the $z$-direction.} the metric reads
\begin{eqnarray}
\label{eq:metricansantz}
& ds^2=-f(v,u) dv^2 - \frac{2L^2}{u^2} dvdu 
  + \Sigma(v,u)^2 \left[ e^{\xi(v,u)} (dx^2 + dy^2) + e^{-2\xi(v,u)} dz^2 \right],~~ \label{eq:ansatz}
\end{eqnarray}
where $v$ and $u$ denote the time and radial coordinate, respectively. The conformal boundary is located at $u=0$ where we impose that the metric asymptotes to AdS$_5$.
The minimal ingredients to generate the CME are a finite axial charge density $q_5$ and a (vector) magnetic field $B$. Hence, the gauge field ansatz reads (in radial gauge $V_u=A_u=0)\,$ 
\begin{equation}
\begin{split}
&  V = \frac{B}{2} (xdy-ydx) +V_z(v,u)\,dz\,,  \hspace{1.2cm}
  A = -Q_5(v,u)\, dv \,. \label{gauge:A}
\end{split}
\end{equation}
The equations of motion for our ansatz~\eqref{eq:ansatz} and \eqref{gauge:A} may be found in~\cite{Ghosh:2021naw}. The asymptotic expansions read
\begin{align}
&  Q_5(v,u)=\frac{u^2}{2}q_5+\mathcal{O}(u^3)\,,\ \ \, \quad\quad V_z(v,u)=u^2\, V_{2}(v)+ \mathcal{O}(u^3)\,,\\
& \Sigma (v,u)=\frac{1}{u}+\lambda(v)+\mathcal{O}(u^5)\,,\quad\quad \xi(v,u)=u^4 \left( \xi_4(v)-\frac{B^2}{12} \,\log(u) \right)+\mathcal{O}(u^5) \,, \\
& f(v,u)=\left(\frac{1}{u}+\lambda(v)\right)^2 + u^2 \left( f_2+\frac{B^2}{6}\,\log(u) \right) -2 \Dot{\lambda}(v) +\mathcal{O}(u^3).
\end{align}
In the asymptotic expansions, we introduced $\lambda(v)$ which we use to keep the position of the apparent horizon of the black brane at a fixed radial position $u_h=1\,$ throughout the time evolution making use of the diffeomorphism freedom~\cite{Fuini:2015hba,Ghosh:2021naw}. The coefficient $f_2$ is related to the energy density of the black brane and the subleading coefficients $V_2(v)$ and $\xi_4(v)$ shall give us the vector current and the pressure anisotropy, respectively. We can extract the one-point functions using the asympotic expansions of the solution
\begin{equation}
\label{eq:QFTcurrent}
    2\kappa^2\left<J_z\right> = 2V_2(v)\,,\quad \quad 2\kappa^2\left<J_5^0\right> = q_5,
\end{equation}
\begin{equation}
\label{eq:stress}
    \begin{split}
        & \expval{T^v_{\ v}}=\frac{1}{4 \kappa^2}\left[ 6 f_2 -B^2 \textrm{log}(\mu L) \right]\,, \quad\quad  \expval{T^z_{\ z}}=-\frac{1}{4 \kappa^2} \left[ 2f_2+16\xi_4(v)+B^2 \textrm{log}(\mu L) \right],\\
        & \expval{T^x_{\ x}}=\expval{T^y_{\ y}}=-\frac{1}{8\kappa^2}\left[B^2+4 f_2-16\xi_4(v) -2 B^2\textrm{log}(\mu L) \right]\,,
    \end{split}
\end{equation}

\noindent
for the currents and the stress tensor. The presence of the magnetic field induces a trace anomaly thus breaking conformal invariance at the microscopic level. Hence, we have re-instated the AdS radius $L$ in the expressions for the one-point functions since the regularization procedure depends on a renormalization energy scale $\mu$ due to the broken conformal invariance. All in all, the problem reduces to solving the full dynamics in the bulk, finding the subleading coefficients of $\xi$ and $V_z$ and substituting them into \eqref{eq:QFTcurrent} and \eqref{eq:stress}. The details about the numerical implementation can be found in \cite{Ghosh:2021naw}. Our numerical code is implemented in the programming language \textit{Julia}~\cite{bezanson2015julia}.

The thermodynamic properties of the system are specified by two quantities: temperature and axial chemical potential. At late times, the system equilibrates and we shall label different solutions in terms of their thermodynamical variables in the final equilibrium state. The thermodynamic parameters are also useful to interpret our solutions in terms of results for the QGP produced in heavy-ion collisions. In the dual gravity picture, equilibration implies that the metric becomes stationary at late times and the temperature is given by the black brane 
$T=-u_h^2/(4\pi) f'(v\to\infty,u_h).$
The chemical potential is the difference of $Q_5$ at the boundary and horizon, i.e. $\mu_5 = Q_5(v\to \infty,u_h) - Q_5(v\to \infty,0).$

We conclude this section by discussing the initial state of the dual quantum field theory. The asymptotic form of the metric ansatz \eqref{eq:metricansantz} is dual to an infinite-volume non-expanding plasma. By construction the plasma has a uniform charge distribution given by $q_5\,$, is immersed in a magnetic field of magnitude $B$ and its uniform energy density is $\epsilon\,$. All three of them are considered to be homogeneous and constant in time in our model. Finally, we specify the initial conditions of the evolution by giving an initial profile to the fields $\xi$ and $V_z$. In particular, we choose them to be zero everywhere, which implies that the CME current vanishes in the initial state and the \textit{dynamical} pressure anisotropy, i.e. the anisotropy generated by $\xi_4\,$, is zero. Notice that the initial state is anisotropic since the term proportional to $B$ in \eqref{eq:stress} yields a static contribution to the anisotropy in the pressure (kinematic pressure anisotropy, see \cite{Fuini:2015hba} for details). Hence, for finite magnetic field, the inital state does not describe an equilibrium solution thus driving the out-of-equilibrium dynamics.    
\section{Results}
To gain an intuition, we monitor the features of our model in terms of the CME current $\left<J\right>$ and the dynamical pressure anisotropy $\xi_4$ for different strengths of the magnetic field in section~\ref{sec:bdep}. The parameter scan is performed at fixed energy density $\epsilon_L = 12\,$.\footnote{Working with a different value for $\epsilon_L$ only modifies the final equilibrium state for the pressure anisotropy but does not alter the relevant qualitative behavior like the build up time and the presence or absence of oscillations.} In section \ref{subsec:matched}, we match our model to QCD and give physically relevant values for the parameters.

There is a subtlety related to the definition of the energy density due to
the non-trivial renormalization scale dependence in \eqref{eq:stress}: $\epsilon(\mu) = 2 \kappa^2 \langle T^{vv}\rangle,$ which has to be fixed. 
In this subsection, we simply choose $\mu=1/L$ for computational convenience. Note that the choice $\mu=1/L$ is not physical since the scale can be changed without changing the values of the physical observables at the boundary due to a scaling symmetry~\cite{Fuini:2015hba}. 
The physically relevant scale is $\mu = \sqrt{B}$, which we use in section~\ref{subsec:matched}. Both choices are related through
\begin{equation}
    \dfrac{\epsilon_B}{B^2}=\dfrac{\epsilon_L}{B^2} + \dfrac{1}{4}  \textrm{log}(B L^2)\,,\label{eq:relebel}
\end{equation}\noindent
where $\epsilon_L$ and $\epsilon_B$ refer to the energy density at scales $\mu=1/L$ and $\mu = \sqrt{B}$, respectively. 
\subsection{$B$-dependence}\label{sec:bdep}
\begin{figure}[h!]%
    \centering
    \subfloat{\includegraphics[scale=0.27,clip]{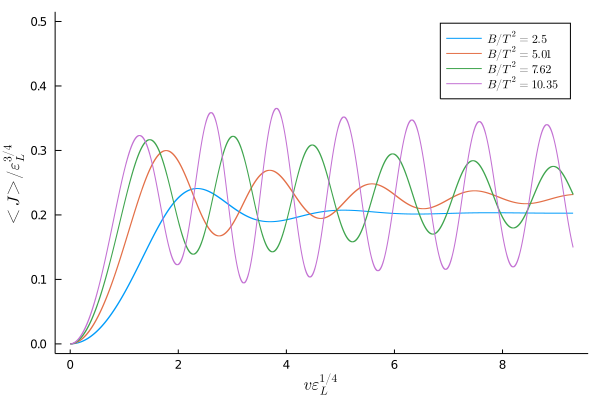}}%
    \qquad
    \subfloat{\includegraphics[scale=0.27,clip]{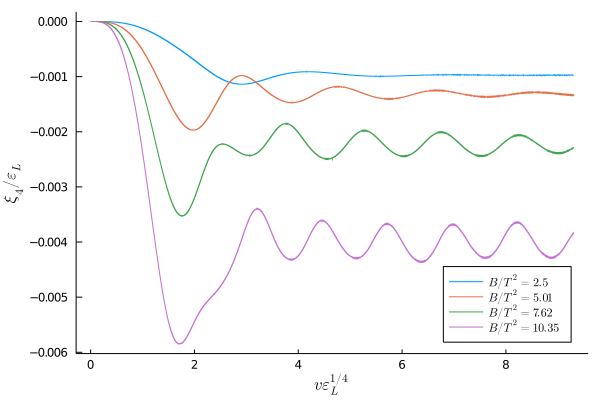}}
    \caption{Vector current (left) and dynamical pressure anisotropy (right
    ) for fixed $\alpha=1.5$ and $q_5=1.5\,$. The magnetic field $B$ is $\{0.5\,,1.0\,,1.5\,,2.0\}\,$. The axial chemical potential of the would-be final equilibrium state is $\mu_5/T=\{0.11\,,0.06\,,0.04\,,0.02\}$.}
    \label{fig:2}
\end{figure}

In this section, we fix the strength of the anomaly $\alpha$ and consider two qualitatively different values of $q_5$ for varying strength of $B$. The results for the larger $q_5$ are shown in figure~\ref{fig:2}. We choose to work with dimensionless variables: time, pressure and current are normalized to the energy density $\epsilon_L$, whereas we consider the dimensionless ratio of $B$ to temperature squared. All thermodynamic quantities refer to the final equilibrium state where they are well defined.

As we increase the magnetic field, we observe the appearance of oscillations in the current $\left<J\right>\,$ in agreement with the quasinormal modes computation in~\cite{Ammon:2016fru,Grieninger:2017jxz}.
As a consequence, the equilibration time increases dramatically for increasing $B$. The oscillatory behavior of the current indicates that the time-evolution is dominated by the lowest QNM near the real axis. The final equilibrium value matches the equilibrium expression for the CME, i.e. $2\kappa^2 \left<J\right>_{eq} = 8\alpha \mu_5 B\,$. Note that oscillatory behavior indicates that we have not reached the final equilibrium state yet.
However, we verified that the axial chemical potential $\mu_5$ read off from these states matches the values of the would-be equilibrium state closely.

It is also worth noting that the current builds up faster for larger $B$. 
Such behavior is expected in $1+1$ dimensional systems~\cite{Ghosh:2021naw}.
For the chosen $q_5$, we clearly observe oscillatory behavior in $\xi_4$ which is not present for small $q_5$ (see fig. 1 in \cite{Ghosh:2021naw}).\footnote{At small $q_5$ oscillations are still there but their amplitude is significantly smaller than the final equilibrium value.}  
\subsection{Matching to QCD}\label{subsec:matched}
\subsection*{Parameters}
In this section, we aim to provide simulations in the parameter range experimentally relevant for the Quark-Gluon plasma (QGP). We obtain estimates for the parameters in our model by matching to known QCD results, i.e. the entropy and the anomaly. Under an axial gauge transformation, our action has the mixed anomaly $\mathcal{A}_{CS} = \frac{\alpha}{2 \kappa^2}$. In order to get an estimate for $\kappa$ we take the entropy of a black brane $ s_{BH} = \frac{A}{4G_N} = \frac{4\pi^4T^3}{2 \kappa^2}\,.$

In the following, we match these gravity expressions to the entropy of QCD at finite temperature and to the axial anomaly~\cite{Ghosh:2021naw,Grieninger:2021rxd}. First, we fix the number of flavors. The up and down quarks are light, whereas the strange quark has a mass of around
$95$ MeV. The cross over temperature of QCD is at around $175$ MeV. Therefore, we include the strange quark in our counting, i.e.
we match to three flavor QCD. The Stefan-Boltzmann value of the entropy density is $s_\text{SB} =  4 \left(\nu_b + \frac{7}{4} \nu_f\right) \frac{\pi^2 T^3}{90},$ where $\nu_b = 2(N_c^2-1)$ and $\nu_f = 2 N_c N_f$ with $N_c=3$ and $N_f=3$. Note that the Stefan-Boltzmann value of the entropy is only reached at asymptotically high temperatures and $s$ is usually smaller at the temperatures of interest. As a ballpark value, we take a factor of $3/4$, which yields $\kappa^2 = (24 \pi^2)/19 \approx 12.5$. Such a relative factor arises
in the strongly coupled $\mathcal N=4$ SYM theory \cite{Gubser:1996de}. QCD lattice simulations also indicate a reduction by a factor of around $0.8$ at
moderate temperatures (see e.g.  \cite{Borsanyi:2013bia}).

The axial anomaly of three flavor QCD is $\mathcal{A}_{QCD} = 2\,\frac{N_c}{32\pi^2}  \left(\frac 4 9 + \frac 1 9 +\frac 1 9\right) = \frac{1}{8\pi^2}\,,$where the factor $2$ comes from the summation over right- and left-handed fermions and in the bracket we sum over the squares of the electric charges of up, down and strange quarks. 
The Chern-Simons coupling $\alpha$ is then determined by matching $\mathcal{A}_{CS} = \mathcal{A}_{QCD}$ leading to $\alpha \approx 0.316\,$.

In typical nucleations of the QGP the temperature, magnetic field and chemical potential $(T,\mu_5,B)$ in [$\text{M}\si{\eV}$] are  $(300,10,m_\pi^2)$ (RHIC) and $(1000,10,15m_\pi^2)$ (LHC). The parameter estimates should be viewed as ballpark values representative for RHIC and LHC physics since there are considerable uncertainties in the values. 
The estimates provide us with two independent dimensionless quantities, which have to be adjusted in the numerical simulations with our two free parameters $(\epsilon_L, q_5)$. The dimensionless ratio $\epsilon_B/B^2$ uniquely fixes $B/T^2$ and hence we work with $\epsilon_B$ in this section. In contrast to the previous section, we plot the full pressure anisotropy evaluated at the physical renormalization scale $\mu=\sqrt{B}$ in this section
\begin{equation}
  \delta P_i \equiv 2\kappa^2\,\frac{\Delta P_B}{B^2}=12\,\frac{\xi_4(v)}{B^2}+\frac 12\,\log(BL^2)-\frac{1}{4}.\label{eq:relpressure}
\end{equation}
Fixing $\epsilon_B/B^2$ in eq. \eqref{eq:relebel} does not fix $f_2$ and $B$ uniquely but rather gives us $B(f_2)$. This means that at fixed $\epsilon_B/B^2$ and vanishing initial dynamical anisotropy $\xi_4(0)=0$, we are confronted with a one parameter family of relative pressures of the initial state~\eqref{eq:relpressure} depending on the value of the magnetic field $B$ (for $L=1$). We shall exploit this feature to study equilibration for several non-equivalent initial states by considering different values of $\delta P_i$.

\subsection*{Simulation}
\begin{figure}[h!]%
    \centering
    \subfloat{\includegraphics[scale=0.27,clip]{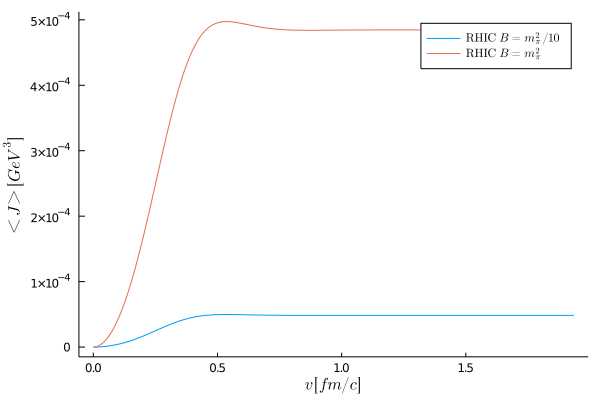}}%
    \qquad
    \subfloat{\includegraphics[scale=0.27,clip]{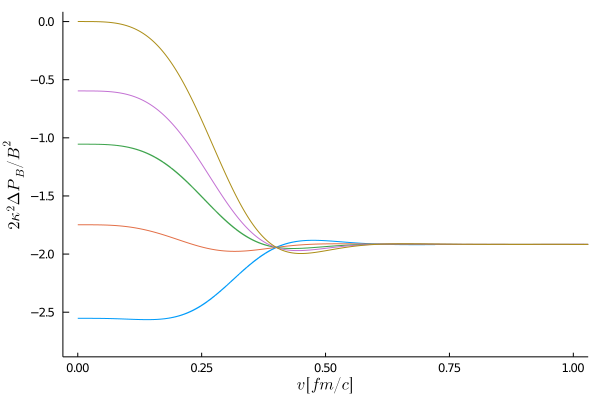}}
    \caption{Vector current (left) and pressure anisotropy (right) as a function of time for the physical parameter estimates for RHIC and anomaly $\alpha\simeq 0.316$; for $m_\pi=140\,\text{M}\text{eV}$. The pressure anisotropy is for $B=m_{\pi}^2$, the results for $B=0.1\,m_{\pi}^2$ are qualitatively similar.}
    \label{fig:8}
    
\end{figure}

\begin{figure}[h!]%
    \centering
    \subfloat{\includegraphics[scale=0.27,clip]{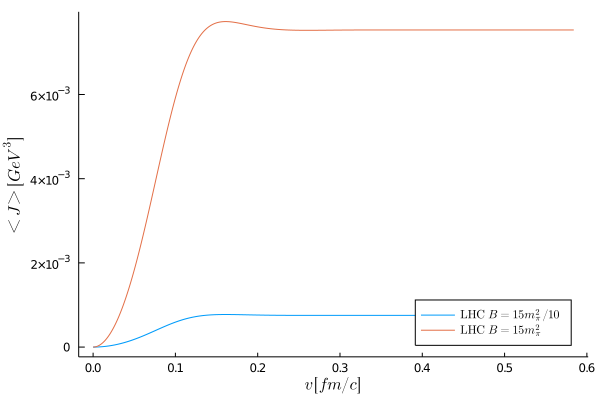}}%
    \qquad
    \subfloat{\includegraphics[scale=0.27,clip]{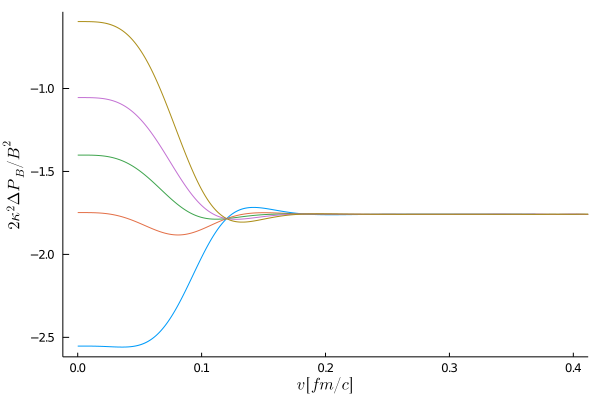}}
    \caption{Vector current (left) and pressure anisotropy (right) for the parameter estimates for LHC ($\alpha\simeq 0.316\,$).  $B=15m_{\pi}^2$ in the right plot; results for $B=1.5m_{\pi}^2$ are qualitatively similar.}
    \label{fig:9}
    
\end{figure}

In figures \ref{fig:8}-\ref{fig:9}, we present the results for the pressure and the current for RHIC and LHC with the physical parameters estimated in the previous section. We fix our initial state by setting the dynamical pressure anisotropy to zero, fix the ratio $\epsilon_B/B^2$ and $q_5$ so that we reach the desired final equilibrium state.
Neither the vector current nor the pressure anisotropy show oscillatory behavior. The former takes slightly more time to build up than the latter. We observe that in the RHIC simulation in figure~\ref{fig:8}, the peak in the vector current is reached at $v_{peak}\sim 0.54 \,$fm/c and the pressure anisotropy reaches its peak at $v_{peak}\sim 0.48 \,$fm/c. 

We display the equilibration times for the simulation with the RHIC parameters in table~\ref{tab:RHICeq}. We use the definition of Chesler and Yaffe to define the equilibration time~\cite{Chesler:2008hg}, i.e. the time where the observable is within 10\% of its final value. As for the LHC in figure~\ref{fig:9}, we find the peak in the vector current at $v_{peak}\sim 0.14 \,$fm/c which is also the time where the pressure anisotropy reaches its peak. We tabulate the equilibration times for the simulation with the LHC parameters in table~\ref{tab:LHCeq}. Note that the equilibration times for the LHC parameters are about 1/3 shorter than in the RHIC case.

A formula to estimate the lifetime of the magnetic field was recently established in~\cite{Guo:2019joy} as $\tau_B \sim 115\,\, \textrm{G}\si{\eV}\textrm{fm/c}/\sqrt{s}$, where $\sqrt{s}$ is the collision energy. At RHIC and LHC the collisions take place at around $\sqrt{s}\simeq 200 \,\textrm{G}\si{\eV}$ and $\sqrt{s}\simeq 5000 \,\textrm{G}\si{\eV}$, respectively, which yield lifetimes of $\tau_B^\text{RHIC}\sim 0.6 \, \textrm{fm/c}$ and $\tau_B^\text{LHC}\sim 0.02 \, \textrm{fm/c}\,$. 
In this context the equilibration and build up times extracted from our simulations may function as a gauge whether the CME is an observable or not.
Our simulations for the RHIC parameters show that the current reaches its equilibrium value before the magnetic field vanishes. On the contrary, for the LHC parameters the magnetic field is too short lived for the current to build up significantly. Hence, we conclude that the chiral magnetic effect should only be observable at RHIC and not at LHC.

In heavy-ion collisions, the magnetic field drops almost instantaneously from its peak value to a base value which is approximately 10\% of the original strength at which it stays for most of its remaining lifetime. Since we approximate the magnetic field as static, we did a second simulation for our parameter estimates with 10\% of the peak magnetic field. The resulting equilibration times remain effectively unchanged and do not affect the previous results on the observability of the CME at either accelerator. We conclude by remarking that the estimate for the axial chemical potential is the most uncertain one in the literature. To prove that our build up and equilibration times are not influenced by our particular choice of the axial chemical potential $\mu_5=10\textrm{M}\si{\eV}$, we performed an analogous simulations at a ten times larger axial chemical potential of $\mu_5=100\textrm{M}\si{\eV}$ \cite{Ghosh:2021naw}. The bottom line is that our results for the build up times and thus the presence of the CME in heavy-ion collisions at RHIC and LHC remain qualitatively unchanged for the larger axial chemical potential.


\begin{table}[h!]
\begin{center}
\begin{tabular}{l c c c c c}
\toprule
RHIC $B=m_\pi^2$ &&&&&  		\\ \addlinespace
\midrule
$\delta P_i$ &-2.55&-1.75&-1.05&-0.60&0.00  \\
$v^{\langle J\rangle}_\text{eq}$ \hspace{0.22cm}in [fm/c]		 &0.380&0.380&0.380&0.380&0.380  \\
$v^{\langle \Delta P\rangle}_\text{eq}$ in [fm/c]		 &0.383&0.418&0.334&0.344&0.350 \\\addlinespace
\toprule
RHIC $B=0.1m_\pi^2$ &&&&&  		\\ \addlinespace
\midrule
$\delta P_i$ &-3.70&-2.90&-2.55&-2.21&-1.75  \\
$v^{\langle J\rangle}_\text{eq}$ \hspace{0.22cm}in [fm/c]		 &0.380&0.380&0.380&0.380&0.380  \\
$v^{\langle \Delta P\rangle}_\text{eq}$ in [fm/c]		 &0.383&0.418&0.310&0.334&0.344 \\
\bottomrule
\end{tabular}
\caption{Equilibration times $v_\text{eq}$ for the RHIC simulation at $B=m_\pi^2$ and $B=0.1\,m_\pi^2$; $\delta P_i$ labels the different initial conditions for the pressure anisotropy~\eqref{eq:relpressure}.}
\label{tab:RHICeq}
\end{center}
\end{table}
\begin{table}[h!]
\begin{center}
\begin{tabular}{l c c c c c}
\toprule
LHC $B=15\,m_\pi^2$ &&&&&  		\\ \addlinespace
\midrule
$\delta P_i$ &-2.55&-1.75&-1.40 &-1.05&-0.60 \\
$v^{\langle J\rangle}_\text{eq}$ \hspace{0.22cm}in [fm/c]		 &0.114&0.114&0.114&0.114&0.114  \\
$v^{\langle \Delta P\rangle}_\text{eq}$ in [fm/c]		 &0.114&0.187&0.085&0.098&0.103 \\\addlinespace
\toprule
LHC $B=1.5\,m_\pi^2$ &&&&&  		\\ \addlinespace
\midrule
$\delta P_i$ &-3.70&-2.90&-2.55&-2.21&-1.75  \\
$v^{\langle J\rangle}_\text{eq}$ \hspace{0.22cm}in [fm/c]		 &0.114&0.114&0.114&0.114&0.114  \\
$v^{\langle \Delta P\rangle}_\text{eq}$ in [fm/c]		 &0.114&0.187&0.085&0.098&0.103 \\
\bottomrule
\end{tabular}
\caption{Equilibration times for the LHC simulation at $B=15\,m_\pi^2$ and $B=1.5\,m_\pi^2$; $\delta P_i$ labels the different initial conditions for the pressure anisotropy~\eqref{eq:relpressure}.}
\label{tab:LHCeq}
\end{center}
\end{table}

\section{Conclusions}
In this work, we investigated the out-of-equilibrium behavior of the chiral magnetic effect in the presence of strong external magnetic fields. We characterize how the chiral anomaly, the magnetic field and the axial charge density influence the non-equilibrium response of the chiral magnetic vector current and the pressure anisotropy and how they affect their equilibration and build up times. We furthermore provide insights on the build up time of the chiral magnetic current in heavy ion collision experiments at RHIC and LHC. Within our setup, the build up time of the chiral magnetic effect is smaller than the lifetime of the magnetic field and should thus be an observable in heavy ion collisions at RHIC~\cite{Shi:2019wzi}. However, the lifetime of the magnetic field at LHC seems to be so short that the magnetic field already drops to zero before the chiral magnetic current can build up in a meaningful way. 
Interestingly, for RHIC we find a shorter equilibration time of $\sim0.35$ fm/c (for an initial state with $\delta P_i(0)=0$) in presence of the chiral anomaly compared to the result of Chesler and Yaffe which estimates the equilibration time as $\sim0.5$ fm/c~\cite{Chesler:2008hg}. This is in agreement with the equilibration time estimate of ~$\sim 0.3$ fm/c for plasma temperatures of $T\sim300-400$ MeV~\cite{Heinz:2004pj}.

\small
\noindent\textbf{Acknowledgments:} SG would like to thank the organizers of the (virtual) Quark Confinement and the Hadron Spectrum 2021 conference in Stavanger, Norway, for the opportunity to present this talk. The authors thank Jewel Gosh and Karl Landsteiner for fruitful collaboration in~\cite{Ghosh:2021naw}. SG is supported by the `Atracci\'on de Talento' program (2017-T1/TIC-5258, Comunidad de Madrid) and through grants SEV-2016-0597 and PGC2018-095976-B-C21. SMT is supported by an FPI-UAM predoctoral fellowship.\vspace{-0.15cm}
\bibliography{HoloCME}
\end{document}